# Spectral engineering of optical microresonators in anisotropic lithium niobate crystal

*Ke Zhang†, \*, Yikun Chen†, Wenzhao Sun, Zhaoxi Chen, Hanke Feng, Cheng Wang\**

*†*These authors contributed equally to this work.

K. Zhang, Y. Chen, Z. Chen, H. Feng, C. Wang

Department of Electrical Engineering & State Key Laboratory of Terahertz and Millimeter Waves, City University of Hong Kong, Kowloon, Hong Kong, China

E-mail: kzhang54@cityu.edu.hk (K. Zhang); cwang257@cityu.edu.hk (C. Wang)

W. Sun

City University of Hong Kong (Dongguan), Dongguan, China

Centre of Information and Communication Technology, City University of Hong Kong Shenzhen Research Institute, Shenzhen, China

Department of Electrical Engineering & State Key Laboratory of Terahertz and Millimeter Waves, City University of Hong Kong, Kowloon, Hong Kong, China

On-chip optical microresonators are essential building blocks in integrated optics. The ability to arbitrarily engineer their resonant frequencies is crucial for exploring novel physics in synthetic frequency dimensions and practical applications like nonlinear optical parametric processes and dispersion-engineered frequency comb generation. Photonic crystal ring (PhCR) resonators are a versatile tool for such arbitrary frequency engineering, by controllably creating mode splitting at selected resonances. To date, these PhCRs have mostly been demonstrated in isotropic photonic materials, while such engineering could be significantly more complicated in anisotropic platforms that often offer more fruitful optical properties. Here, we realize the spectral engineering of chip-scale optical microresonators in the anisotropic lithium niobate (LN) crystal by a gradient design that precisely compensates for variations in both refractive index and perturbation strength. We experimentally demonstrate controllable frequency splitting at single and multiple selected resonances in LN PhCR resonators with different sizes, while maintaining high $Q$-factors up to $1 \times 10^6$. Moreover, we experimentally construct a sharp boundary in the synthetic frequency dimension based on an actively modulated x-cut LN gradient-PhCR, opening up new paths toward the arbitrary control of electro-optic comb spectral shapes and exploration of novel physics in the frequency degree of freedom.

## 1. Introduction

Optical micro-resonators, including micro-ring resonators, micro-disk resonators, and photonic crystal cavities, play an unrivaled role in science and engineering. These resonant devices, with dramatically enhanced light-matter interactions, have found numerous applications in optical sensing[1], nonlinear optics[2], quantum optics[3-5], optomechanics[6-7], and synthetic frequency dimension physics[8]. Due to the whispering-gallery-mode (WGM) nature, micro-ring/disk resonators typically feature equally or nearly equally spaced resonances, which on the one hand is the foundation for applications like frequency comb generation[2, 9-10]. On the other hand, however, it is usually difficult to independently control the precise locations of these resonances much beyond a near-periodic distribution. Such control capability could allow the resonator dispersion profiles to reach a much broader parameter space, enabling applications like broadband nonlinear optical parametric processes[11-12], arbitrary spectral shaping of frequency combs[13-14], and construction of complex synthetic frequency crystals[15-16].

One powerful solution to locally engineer the resonant frequencies is by inducing controlled mode splitting at selected resonances in a photonic crystal ring (PhCR) resonator[12-14, 17-18]. By carefully designing the periodic structural modulation (i.e., photonic crystal) along the microring circumference, controllable coupling between the clockwise (CW) and counter-clockwise (CCW) modes will occur, leading to normal mode splitting at the target resonant frequencies only, while the other resonances remain unchanged[12]. This selective frequency-engineering tool has been successfully demonstrated in a number of isotropic photonic material platforms, e.g., silicon nitride ($Si_3N_4$)[12, 17] and tantalum pentoxide ($Ta_2O_5$)[18], and applied for advanced dispersion engineering and comb spectral shaping[13-14].

However, achieving such independent resonance engineering is much more complicated in anisotropic crystals, where the refractive indices vary along different crystal directions. In this case, the optimal structural modulation period for a particular optical mode also changes along the ring circumference. As a result, a single-period PhCR design will lead to phase-matched mode splitting at multiple resonant modes, significantly complicating the PhCR designs especially in applications requiring precise spectral engineering. On the other hand, realizing such anisotropic PhCRs in a controllable manner could substantially broaden the application scope of these systems, since the crystal anisotropicity is often closely associated with material properties not available in isotropic materials. One such prominent example is lithium niobate (LiNbO$_3$, LN), which features a large electro-optic (EO) coefficient ($r_{33}$ ~31 pm/V) that is only existent in non-centrosymmetric crystals, in addition to good linear optical properties, such as a moderately high refractive index (~2.2) and a wide optical transparency window (0.4 - 5 μm)[19-20]. The last decade has witnessed the rapid development of integrated photonic devices based on the LN-on-insulator (LNOI) platform[21-32]. Leveraging high-$Q$ LNOI micro-resonators and the excellent electro- and nonlinear-optic properties of LN, numerous devices with unprecedented performances have been demonstrated, e.g., broadband optical frequency comb sources[21-26], quantum microwave-to-photon converters[27-28], unidirectional frequency shifters and beam splitters[29], as well as ultra-efficient nonlinear wavelength converters[30-31]. Most of these functional devices have been achieved in x-cut LN thin films where the largest EO tensor component $r_{33}$ (corresponding to $\chi^{(2)}_{zzz}$) aligns with the transverse-electric (TE) mode and could be conveniently accessed with planar modulation electrodes[24-29]. However, LN is a uniaxial anisotropic material with a substantial difference between the refractive index in the *x-y* crystal plane and that along the *z*-crystal direction ($\Delta n$ ~0.07), leading to difficulties in the precise mode-splitting engineering of LN PhCR resonators due to the aforementioned reasons. In addition, controllable inscription of the nano-scale width-modulation features is also challenging in LN, a well-known difficult-to-etch material[32]. As a result, PhCR resonators with arbitrary resonant frequency engineering capabilities have not been realized in the LNOI platform so far.

In this article, we overcome these limitations and demonstrate the precise spectral engineering of LNOI PhCRs in both z-cut isotropic and x-cut anisotropic LN thin films. By carefully controlling the period and amplitude of the geometric modulation at the inner boundary of the microrings, we achieve mode splitting at single and multiple selected resonances in LN PhCR resonators of different sizes with controllable splitting strengths, while maintaining high $Q$-factors up to $1 \times 10^6$. Most importantly, we implement a novel gradient design based on effective index modulation, which could precisely compensate for the anisotropic nature of x-cut LN PhCR resonators, leading to single mode splitting at arbitrarily chosen resonances without affecting neighboring modes. Using this universal spectral engineering tool, we experimentally demonstrate the construction of a sharp boundary in the synthetic frequency dimension in an anisotropic x-cut LN microresonator. We show effective trapping and reflection of optical power flow at the synthetic frequency mirror with a measured extinction ratio of 26.7 dB between the two sides of the mirror.

## 2. Results and discussions

### 2.1. Working principle of mode engineering in isotropic and anisotropic LN PhCR resonators

**Figure 1a-d** shows a schematic comparison of TE-polarized optical resonances in four scenarios: a simple micro-ring resonator with equally-spaced resonant dips (**a**), uniform PhCR resonators in isotropic (**b**) and anisotropic (**c**) platforms, as well as the proposed gradient PhCR resonators in an anisotropic platform (**d**). As **Figure 1b** shows, in an isotropic system like $z$-cut LN where the refractive index is constant in the $x$-$y$ plane ($n_x = n_y = 2.211$), we introduce mode splitting at a selected resonance by adding a periodic structural modulation (photonic crystal) at the inner boundary of a micro-ring resonator. Specifically, the width of this normal PhCR follows $w = w_0 + \frac{1}{2} \cdot A \cdot \cos(n\phi)$, with $w_0$ the initial resonator width, $A$ the peak-to-peak amplitude of the geometric modulation, $n$ the modulation index, and $\phi$ the ring azimuthal angle. In such isotropic PhCRs, mode splitting appears when and only when the azimuthal mode number $m$ is phase-matched with the geometric modulation index $n$, i.e., $n = 2m$, while the other resonances ($n \neq 2m$) remain intact[11-12]. Physically, this happens when the structural modulation period is exactly half of the effective wavelength of the selected mode, which is a constant in an isotropic system. In other words, one target mode corresponds to one fixed structural modulation period. However, such uniformly modulated PhCR loses the simple phase-matching relationship in an anisotropic system like x-cut LN ($n_y = 2.211$, $n_z = 2.138$), as **Figure 1c** shows. The varying effective indices (shown as gradient color) at different locations of the ring are phase-matched with resonances at different frequencies for a fixed structural modulation period, resulting in complicated and unwanted multiple splitting events near the target mode. To solve this problem, in this work, we employ PhCR resonators with a gradient design in x-cut anisotropic LN (**Figure 1d**), which effectively compensates for variations in both refractive indices and geometric-perturbation strength, leading to dramatically suppressed mode splitting at un-targeted modes. This novel gradient design is realized by precisely tailoring the sizes of each structural perturbation unit, such that the effective indices at every tip or indent point along the circumference remain constant. This drives the system back to the isotropic PhCR model, resulting in single mode splitting at selected resonance when $n = 2m$ (**Figure 1d-e**). At the target resonances, the degeneracy between initially de-coupled CW and CCW modes are lifted, forming two new standing-wave modes with slightly different eigen frequencies, which correspond to the in-phase and out-of-phase superpositions of CW and CCW waves. Insets of **Figure 1e** illustrate the simulated electric field distributions of the two split modes, where the electrical field maxima are respectively located at the widest/narrowest parts of the PhCR resonator, leading to slightly decreased/increased resonant frequencies of $\omega_m^-$ and $\omega_m^+$, respectively. **Figure 1f** shows representative scanning electron microscope (SEM) images of fabricated z-cut LN PhCR resonators in the waveguide-coupling areas, with various geometric modulation amplitudes *A* ranging from 50 nm to 600 nm as designed.

**2.2. Single-tone geometric modulation in z-cut LN PhCR resonators**

We first investigate the mode splitting phenomena in isotropic z-cut LN PhCR resonators. A single-tone sinusoidal geometric modulation is engineered at the inner boundary of a 1.2-μm-width microring resonator (with a radius of 80 μm), where the modulation amplitude *A* = 150 nm and modulation index *n* = 1200. The corresponding measured optical transmission spectrum for TE mode in **Figure 2a**, with a spectral span >80 nm, shows a single mode splitting at our target mode *m* = 600 at 1570 nm wavelength (shaded area) while the other resonances remain largely unchanged, which is consistent with previous literature[12] as well as our theoretical prediction (see **Supplementary 1**). Importantly, these LN PhCR resonators support high *Q*-factors both at the split and unsplit modes. The two split modes feature high intrinsic *Q*-factors of $9.8 \times 10^5$ and $8.8 \times 10^5$, respectively (**Figure 2b**), whereas the intrinsic *Q* factor of a neighboring un-targeted resonant mode in the same PhCR shows a similar value of $9.3 \times 10^5$ (grey dot in **Figure 2c**). Further varying the modulation amplitude *A* within a relatively large window (from 50 nm to 300 nm) also shows negligible influence on the resonator *Q*-factors, which are consistently at the level of 1 million (**Figure 2c**) and are on par with those of the unperturbed microring resonators (*A* = 0 nm) fabricated from the same chip. This indicates that such geometric modulation strategy does not induce significant extra optical loss and supports high *Q*-factors for both split and unsplit resonant modes, which is essential for applications such as high-resolution sensing/metrology[1], and broadband comb generation[24].

We next show that the splitting amount (defined as the wavelength/frequency difference between the two split modes) can be precisely controlled by fine-tuning the geometric modulation amplitude *A*. **Figure 2d** illustrates the measured splitting amount at the same target mode (*m* = 600) as a function of modulation amplitude *A*, showing a near linear relationship similar to that observed in the previous report in $Si_3N_4$ platform[12]. Besides, a narrower resonator results in a larger splitting amount since optical modes are distributed closer to the waveguide boundary and experience a stronger geometric perturbation effect for the same *A*[33]. The measured splitting amount could reach up to 0.5 nm (62.5 GHz) at *A* = 300 nm (see **Supplementary 2**), sufficient to support applications like dispersion engineering for broadband Kerr comb generation[14]. We also observe parasitic mode splitting at the adjacent modes (*m* =

600±1), as shown by the purple circles in **Figure 2e** with a fitted slope of 0.09 pm/nm. This unwanted splitting is likely caused by slight deviation in fabricated structures from the ideal geometric modulation profile, but is insignificant (<5%) compared with the target mode splitting amount (slope of 2 pm/nm)[12].

We further demonstrate the capability to precisely control the spectral location of mode splitting (or mode number $m$) by changing the structural modulation index $n$. The splitting mode wavelength can be selectively tuned from 1570 nm to 1530 nm (corresponding to azimuthal mode numbers $m_1$ = 600 and $m_2$ = 620), while maintaining constant splitting amounts for specific modulation amplitudes (**Figure 2f**). The flexible control over the splitting location and splitting strength can also be extended to PhCRs with different radii and in turn with different free spectral ranges (FSR). For example, we show that PhCRs with a doubled radius (160 µm here) give rise to nearly the same splitting amounts for the same modulation amplitudes (**Figure 2g**), as long as the structural modulation index $n$ is also doubled (from 1200 to 2400 in this case) to match with the doubled target-mode azimuthal number (from 600 to 1200), as expected from the perturbation theory[12, 33]. The ability to achieve on-demand mode-splitting location and strength in resonators with different FSRs is crucial for applications like arbitrary spectral shaping of Kerr/EO combs with different repetition rates. Besides, the coupling gap between PhCR and the bus waveguide has negligible influence on the mode-splitting location and strength (see **Supplementary 2**), making such PhCRs robust in practical applications requiring different resonator coupling states.

### 2.3. Dual-tone geometric modulation in z-cut LN PhCR resonators

We then show the generation of mode splitting at two arbitrarily chosen azimuthal modes in z-cut PhCR resonators through a dual-tone geometric modulation. Here the PhCR width follows a linear combination of two independent single-tone modulation, in the form of $w = w_0 + \frac{1}{2} \cdot A_1 \cdot \cos(n_1\phi) + \frac{1}{2} \cdot A_2 \cdot \cos(n_2\phi)$, where $w_0$ is chosen to be 2 µm in this section. Each of the structural modulation indices $n_i$ is phase-matched to the respective target mode number $m_i$ ($n_i$ = $2m_i$), similar to the single-tone scenario. For a moderate geometric modulation strength ($A_1$ = $A_2$ = 100 nm), as **Figure 3a** shows, two distinct split modes can be identified at the target mode numbers 600 and 620 (pink shadows), while the other modes remain intact. The splitting amounts at these two modes are also similar (~ 0.1 nm) since our designed geometric modulation amplitudes for the two modes ($A_1$ and $A_2$) are the same.

Interestingly, when the modulation amplitude goes beyond the moderate perturbation regime ($A > 200$ nm in our experiments), we observe the generation of spurious mode splitting at mode numbers $2m_1 - m_2$ and $2m_2 - m_1$, as shown in the optical transmission spectrum in **Figure 3b**. Here we set a more aggressive dual-tone geometric modulation of $A_1 = A_2 = 300$ nm for target mode numbers of $m_1 = 600$ and $m_2 = 610$. The SEM image of this fabricated device is shown in **Figure 3c**. We can clearly see mode splitting not only at the designed target modes 600 and 610, but also at two "sideband" modes 590 and 620 (**Figure 3b**), similar to the four-wave-mixing (FWM) process in nonlinear optics. This spurious mode-splitting phenomenon is likely due to the fabrication process (either during lithography or dry etching) that not only linearly transfers the designed geometric modulation into the final devices, but also introduces nonlinear effects that mix the two spatial modulation frequencies $m_1$ and $m_2$ (**Figure 3d**). It should be noted that the splitting amount of such "sideband" modes (~ 0.1 nm here, grey dot in **Figure 3e**) is significant (~ 19%) compared to the target splitting (~ 0.53 nm, **Figure 3e**), distinguishing themselves from the previously discussed parasitic splitting at adjacent modes (<5%). Further statistical analysis of the splitting amounts as functions of modulation amplitude in the aggressive modulation regime (shaded area in **Figure 3e**) also indicates a significant deviation from the initial linear relationship in the moderate modulation region, which is also observed in the single-tone geometric modulation case (see **Supplementary 2**).

## 2.4. Effective index modulated design for PhCR resonators in anisotropic x-cut LN

Next, we demonstrate controllable engineering of mode splitting in anisotropic x-cut LN PhCR resonators using the proposed gradient design that provides a constant effective index modulation along the PhCR. As discussed in **Section 2.1** and **Figure 1c**, a normal single-tone PhCR in x-cut LN leads to complicated multiple mode-splitting phenomena due to the refractive index variation along different crystal orientations. **Figure 4a** shows such a representative measured optical transmission spectrum for TE mode with multiple split modes as indicated by the shaded areas. The splitting amount at each resonance can be well predicted in theory, by calculating the overlap integral between the fundamental $TE_0$ mode field distribution and the structural modulation profile along the ring circumference[12], in this case with varying LN permittivity term $\varepsilon_{LN}(\phi)$, electric field terms $|E(\phi)|$, and electric displacement terms $|D(\phi)|$:

$$\beta_m = \frac{A\omega_m}{2} \frac{\int dS[(\varepsilon_{LN}(\phi) - 1)\varepsilon_0 |E_\parallel(\phi)|^2 + \left(1 - \frac{1}{\varepsilon_{LN}(\phi)}\right)\frac{1}{\varepsilon_0}|D_\perp(\phi)|^2]\cos^2(m\phi)\cos(n\phi)}{\int dV \varepsilon(\phi)\left(|E_\parallel(\phi)|^2 + |E_\perp(\phi)|^2\right)}$$

where the numerator is a surface integral over the geometric boundary d$S$ that corresponds to the overall perturbation strength at the ring boundary, and the denominator is a volume integral of the total optical energy as a normalization term. $E_\parallel$ and $E_\perp$ ($D_\perp$) are the electric field (electric displacement) components of the optical modes parallel ($\parallel$) and perpendicular ($\perp$) to the geometric boundary d$S$, and $\varepsilon_0$ is the vacuum permittivity. The middle panel of **Figure 4b** summarizes the measured (bars) and simulated (circles) splitting amounts for modes in the vicinity of the target mode ($m = 640$ in this case) for a structural modulation amplitude of $A = 200$ nm. The distribution of spatial Fourier components into several adjacent modes instead of a single mode also substantially reduces the average splitting amount, from ~ 220 pm in z-cut LN PhCR to ~ 20 pm here. In the case of an even smaller modulation amplitude of $A = 50$ nm, mode-splitting phenomena become too weak to be observed in our system (bottom panel of **Figure 4b**). When further increasing the modulation amplitude to 300 nm, we see an elevated overall mode-splitting profile that can still be matched well with our theoretical prediction (top panel of **Figure 4b**). Similar complex multiple-mode-splitting phenomena are also observed for dual-tone modulated PhCRs in x-cut LN, with a number of split modes spreading around each of the target modes (see **Supplementary 3**). These undesired mode splitting events could significantly limit the further applications of PhCR in anisotropic x-cut LN and other anisotropic material platforms.

To address this issue, we implement a gradient PhCR design that successfully suppresses the unwanted mode splitting events and realizes single mode splitting only at the selected mode without affecting other resonances. Our gradient design strategy mainly consists of two aspects. First, we gradually taper the mean width of the PhCR to maintain a constant effective index at different azimuthal angles of the ring, such that a single geometric modulation period only excites a single target mode regardless of the crystal orientation. **Figure 4c** shows our simulated waveguide effective indices for the fundamental $TE_0$ mode as functions of width and azimuthal angle (with respect to crystal *y*-axis). A narrower (wider) waveguide is therefore generally required along the *z*-crystal (*y*-crystal) direction to keep a constant effective index. We note though, only implementing such a varying mean ring width is not sufficient to achieve the desired single mode splitting, since a constant geometric modulation amplitude (as we did earlier) in this case would correspond to different effective perturbation strengths at different locations. For example, a wider waveguide sees less perturbation effect even though the geometric modulation amplitude is the same, since the optical mode is better confined and interacts less with the side surfaces. As a result, the effective modulation (perturbation) profile features a varying envelope with a period of 2 on top of the intended modulation period ($n = 1280$ in this case), which again leads to multiple mode-splitting events (Fourier components at $n$, $n \pm 2$, etc.). To address this issue, we introduce the second compensation strategy that makes use of a varying geometric modulation amplitude profile, such that the effective perturbation strength also stays constant along the ring. This is achieved by carefully tailoring the sizes of each gear-like structure, such that the effective indices at every tip or indent point remain constant. In our final design, the widths at each tip (indent) point along the resonator circumference are chosen to follow the red (blue) curve in **Figure 4c** that features a constant effective index value of ~1.897 (1.891). Such design ensures that not only the mean effective index, but also the tip-indent index difference stays constant throughout the ring.

**Figure 4d** illustrates the measured optical transmission spectrum of our gradient-design x-cut LN PhCR resonator, clearly showing a single mode splitting of ~60 pm at our target mode 640 and proving our compensation strategy effective. **Figure 4e-h** shows the SEM images of the fabricated gradient x-cut PhCR resonator, with zoom-in views of the structural modulation profiles along various crystal orientations. Going from the *y*-propagating (**Figure 4f**) to the *z*-propagating location (**Figure 4h**), our gradient PhCR features a tapering waveguide width from 3 μm to 0.75 μm and a weakening geometric modulation amplitude from 800 nm to 50 nm, yet keeping both the mean effective index and the effective perturbation strength constant. We note that such a substantial mode splitting is enabled only by a small effective index difference (~0.006) between the tip and indent areas, which further verifies that the significant refractive index anisotropy (~0.07) in x-cut LN cannot be neglected. Our gradient-design concept in the anisotropic x-cut LN platform not only proves to be effective in precisely controlling the location and strength of mode splitting, but also can be readily applied to other anisotropic crystal platforms like silicon carbide (SiC), substantially broadening the application scope of PhCR systems.

**2.5. Application: PhCR-induced boundary in the synthetic frequency dimension**

To showcase the application prospect of our anisotropic x-cut LN PhCR system, we further realize the construction of a sharp boundary in the synthetic frequency dimension by integrating EO modulation electrodes to our gradient PhCR. The concept of synthetic frequency dimension makes use of the frequency degree of freedom to explore high-dimensional classical/quantum dynamics on low-dimensional geometries. It has been used to investigate various physical effects, such as topological windings/braiding in non-Hermitian bands[34-35], high-dimensional frequency conversion[36], Bloch oscillations[37] and so on. LN EO microresonators are a promising platform for synthetic frequency dimension physics, since high-$Q$ microresonators provide distinct eigenstates in the frequency domain, while an EO modulation process strongly couples the neighboring states[16, 38]. Frequency-domain boundaries, a fundamental phenomenon in analogy to mirrors in real space[15], have previously been demonstrated in EO modulated LN resonators by mode crossing or by coupling with an auxiliary ring[16]. However, the former approach relies on accidental mode splitting events and lacks precise control over individual resonances, whereas the latter requires multiple resonators with significantly more complicated fabrication and thermal-tuning processes. Here, we overcome these limitations and experimentally demonstrate the deterministic formation of a PhCR-induced frequency boundary.

**Figure 5a-b** shows the microscope and SEM images of our fabricated x-cut LN PhCR, which includes a gradient-designed racetrack resonator and ground-signal-ground (GSG) modulation electrodes. Here a racetrack resonator is used instead of circular rings to maximize the $y$-propagating section, which aligns with the largest EO tensor component $r_{33}$, nevertheless the gradient design rules remain the same as those described above. **Figure 5c** shows the zoom-in SEM image of our gradient PhCR before $SiO_2$ cladding, featuring precisely tailored widths and geometric modulation strengths at each tip (indent) point along the resonator circumference. The resonator is designed for a constant effective index of ~1.918 (1.913) at the tip (indent) point, slightly higher than that in **Section 2.4** due to the additional $SiO_2$ cladding layer (see **Supplementary 4**), which also indicates that our effective-index-modulated gradient PhCR strategy is robust in different scenarios.

Using the EO-modulated PhCR resonator, we realize the deterministic construction of frequency boundary in the synthetic dimension, where the optical modes coupled via EO interactions serve as lattice points of a one-dimensional frequency crystal. **Figure 5d-e** illustrates the measured optical transmission spectra (without active EO modulation) from a reference racetrack resonator-based EO comb generator and our PhCR-based device, respectively. For a fair comparison, the two devices are fabricated on the same chip, with the same resonator FSR (around 25 GHz), electrode gap (8.5 μm) and similar loaded $Q$-factors (2.8 × 10$^5$ and 2.5 × 10$^5$ for the reference and our proposed device, as shown respectively in the insets of **Figure 5d-e**). When setting the input laser at one of the optical resonances and applying a 24.96-GHz radio-frequency (RF) signal to the modulation electrodes matching the FSR of the resonator, efficient optical energy flow between neighboring lattice points of the one-dimensional frequency crystal could be induced, resulting in a typical EO frequency comb spectrum measured at the output port (**Figure 5f**). The measured spectrum features a roll-off slope of 2 dB per lattice point away from the pump mode due to optical energy decay during propagation in the frequency crystal. In contrast to the reference racetrack resonator where all optical modes are equally spaced, the PhCR design induces a 6.25-GHz mode splitting at a
10

target TE mode $m = 6000$, corresponding to a 25% perturbation to the resonator FSR (**Figure 5e**), which serves as a sharp boundary in the frequency dimension. In this case, the normal optical energy flow in the synthetic frequency dimension is interrupted and reflected when meeting this boundary, leading to trapped state of optical energy with constructive/destructive interference pattern at every other lattice point (**Figure 5g**). Notably, the observed extinction ratio between optical energies before and after the boundary is as high as 26.7 dB, indicating a strong boundary effect in our system. Importantly, our proposed PhCR-design strategy could be readily extended to realize multiple boundaries with individually addressable positions and strengths, providing a new platform to explore the topological boundary phenomena in the synthetic frequency dimension.

## 3. Conclusions

We have demonstrated, for the first time, a thorough investigation on the spectral shaping of optical resonances in thin-film LN PhCRs. By carefully engineering the geometric modulation profile, we show the controllable generation of mode splitting at single and multiple selected resonances in isotropic z-cut and anisotropic x-cut LN thin films. We propose and experimentally demonstrate a novel and effective gradient PhCR design that compensates for both the variations in effective index and in perturbation strength along an anisotropic ring. As a proof-of-concept demonstration, we experimentally create a PhCR-induced boundary state that strongly modifies the optical energy flow in the synthetic frequency dimension in an EO-modulated LN PhCR resonator. The ability to arbitrarily control the individual spectral properties of optical micro-resonators is of critical importance to both the exploration of fundamental high-dimensional physics and practical nonlinear-optics applications. For example, the confinement of optical energy between frequency boundaries could be used to investigate the discretization of band structure and the interaction of boundaries with one-way chiral modes in a quantum Hall ladder[15]. Advanced dispersion profiles beyond simple polynomial functions, e.g., weak and broadband anomalous dispersion[10], could be achieved for spectral shaping of Kerr/EO combs leveraging the strong second- and third-order optical nonlinearities in LN. This universal spectral engineering tool could also lead to a more flexible design of nonlinear optical processes, e.g., second-harmonic generation and optical parametric oscillation, leveraging the strong $\chi^{(2)}$ optical nonlinearity in LN. These applications are particularly appealing when using the anisotropic x-cut LN PhCRs demonstrated in this work, which makes full use of the strongest $\chi^{(2)}$ tensor component in LN, benefiting applications such as on-chip optical communications[39-40], light detection and ranging (LiDAR)[41-42], as well as classical and quantum optical computing systems[43-44].

## 4. Methods

*Devices Fabrication*: Our devices are fabricated from a commercially available LNOI wafer (NANOLN), which consists of a sub-micron LN thin film, a 2-μm buried SiO$_2$ layer, and a 500-μm silicon substrate. Micro/nano-structures are first patterned by electron-beam lithography (EBL) process using Hydrogen silsesquioxane (HSQ, FOX16) electron-beam resist, which is then transferred to the LN layer using an Ar$^+$ plasma reactive ion etching (RIE) dry etching process. The residual HSQ resist is removed using buffered oxide etch (BOE), leading to LN rib waveguides with ~ 1:1 ratio between the rib height and the un-etched slab thickness. Specifically, the z-cut devices are fabricated from a 400-nm-thick LN film, while the x-cut devices are from a 500-nm-thick LN film. The choice of z-cut and x-cut wafer thickness is based on wafer availability rather than design needs. For the active frequency-boundary chip, metal electrodes are fabricated using a standard sequence of photolithography, thermal evaporation, and lift-off process, resulting in a 600-nm-thick metal layer (580-nm copper and 20-nm gold) with an electrode gap of 8.5 μm. The metal is located 2.2 μm away from the edge of the PhCR resonator, ensuring strong EO phase modulation while not affecting the resonator $Q$ factor. 600-nm-thick SiO$_2$ is deposited by plasma-enhanced chemical vapor deposition (PECVD) as the cladding layer of our devices. Finally, the fabricated chips are cleaved, and the facets are carefully polished for end-fire coupling.

*Simulation:* We simulate the azimuthal mode distribution of the micro-resonators using COMSOL Multiphysics (COMSOL Inc.). The effective index of the LN waveguide is simulated using finite element methods (Ansys, Inc.).

*Devices Measurement:* Continues-wave (CW) light source from a tunable telecom laser (Santec TSL -550) goes into a fiber polarization controller (FPC) to ensure TE mode excitation before coupled into our devices using a lensed fiber. Another lensed fiber is used to collect the output signal of our devices, which is then sent to a photodetector (Newport, 1544-A) to monitor the optical power. For the measurement of boundary states in the synthetic frequency dimension, RF signals (Anritsu, MG3697C) are amplified using a medium-power gallium arsenide (GaAs) amplifier (Pasternack, PE15A4021) and delivered to the metal electrodes on chip through a high-speed GSG probe (GGB industries). The generated EO comb spectra are monitored using an optical spectrum analyzer (OSA, Yokogawa AQ6370).

**Supporting Information**

Supporting Information is available from the Wiley Online Library or from the author.

**Acknowledgments**

The authors thank Dr. Wing-Han Wong for her help in device fabrication and Dr. Yaowen Hu for valuable discussions. This work is supported in part by Research Grants Council, University Grants Committee (CityU 11212721, N_CityU113/20), Croucher Foundation (9509005).

Received: ((will be filled in by the editorial staff))
Revised: ((will be filled in by the editorial staff))
Published online: ((will be filled in by the editorial staff))

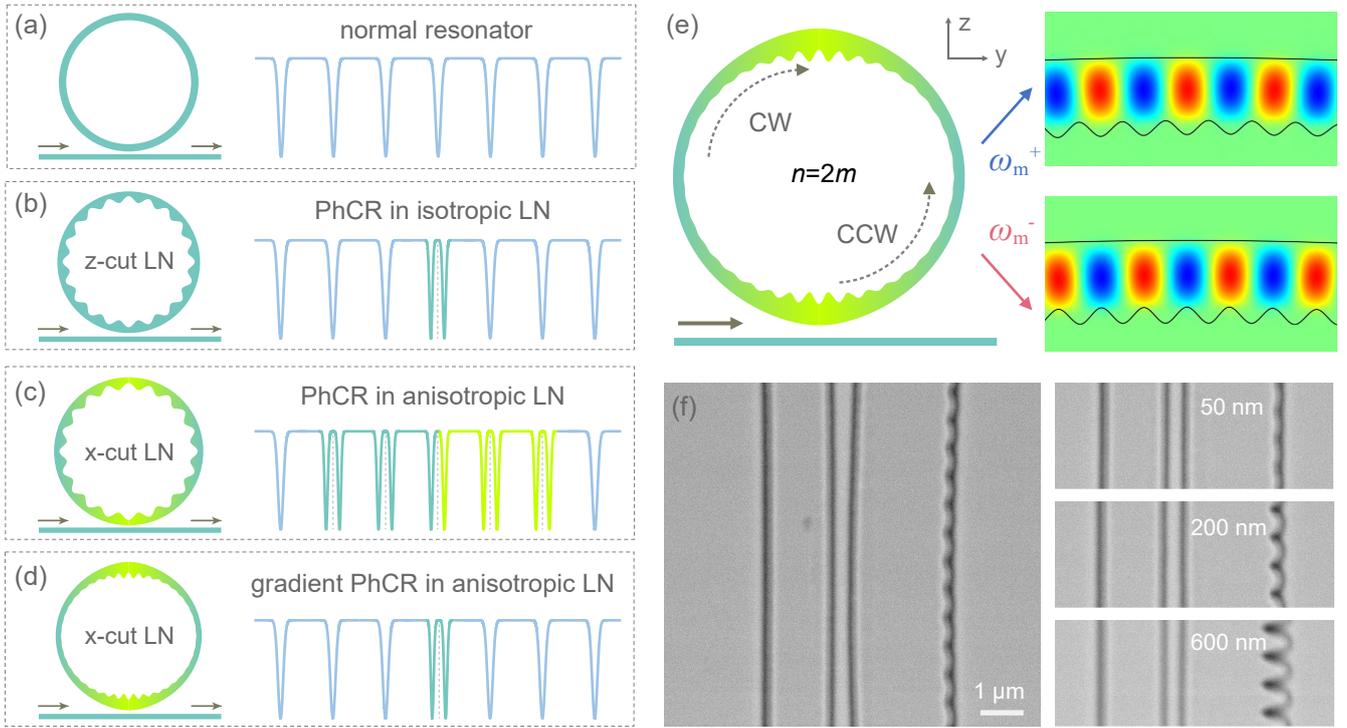

**Figure 1.** Working principle of isotropic and anisotropic LN PhCR resonators. (a) A simple microring resonator features equally spaced resonance dips. (b) A typical PhCR resonator in isotropic z-cut LN with periodic structural modulation, leading to single mode splitting at the target mode. (c) Uniform structural modulation in anisotropic x-cut LN PhCR will lead to multiple split modes originating from different sections of the ring. (d) Our PhCR resonator with gradient design achieves single mode splitting by compensating for the crystal anisotropicity in x-cut LN. (e) The gradient design includes variations in both the center width and the structural modulation strength along the ring circumference, leading to lifted degeneracy between CW and CCW modes and the formation of two new standing-wave modes. Right insets: simulated electric field distributions of the two split modes. (f) SEM images of fabricated LN PhCR resonators in the coupling regions with various geometric modulation amplitudes of 50 nm, 200 nm, and 600 nm.

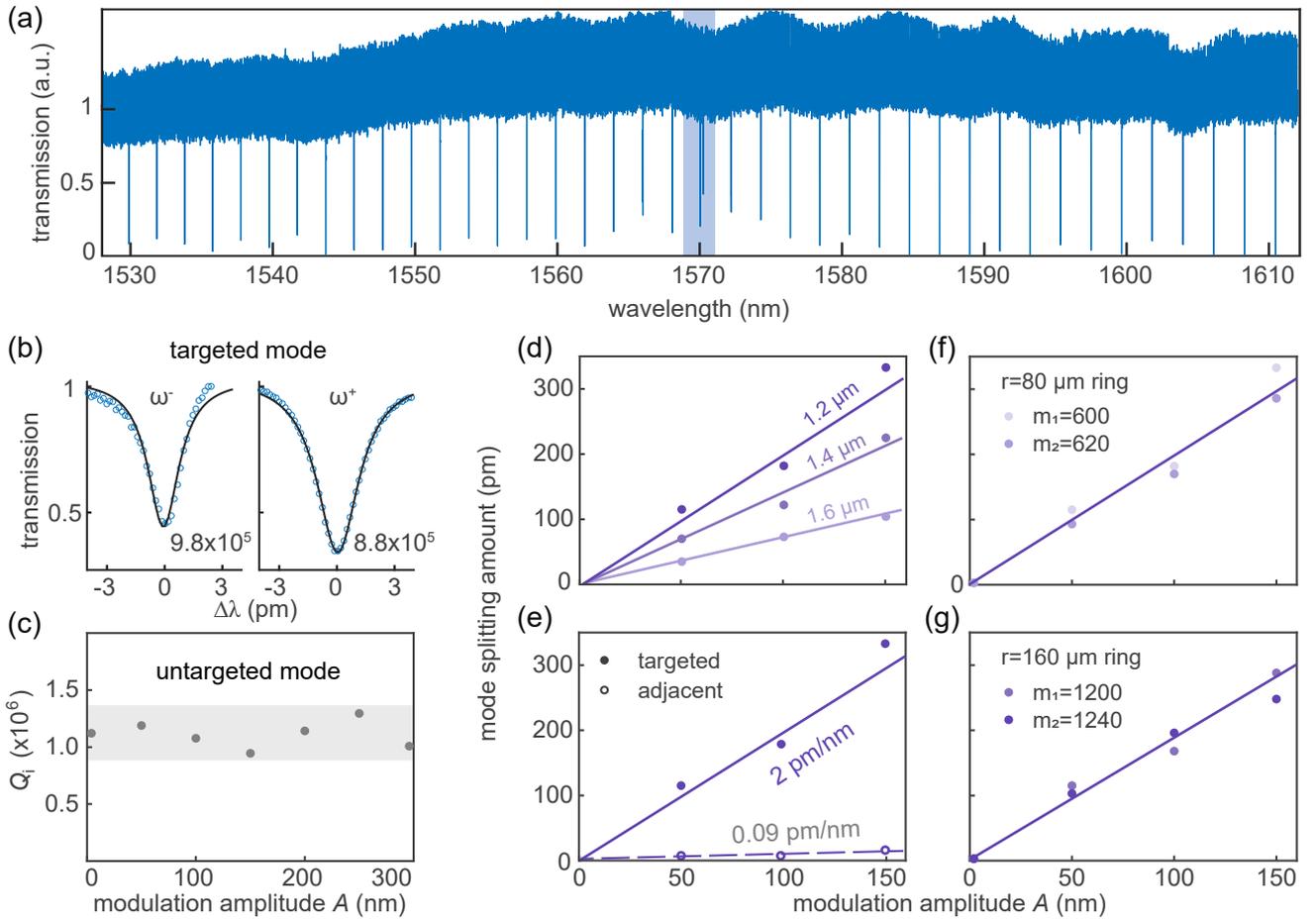

**Figure 2.** Single-tone geometric modulation in z-cut LN PhCR resonators. (a) Measured transmission spectrum of a single-tone modulated z-cut LN PhCR resonator, showing a single mode splitting at 1570 nm wavelength (shaded area). (b) Zoom-in view of the two split resonances with Lorentzian fits indicating intrinsic $Q$-factors of $9.8 \times 10^5$ and $8.8 \times 10^5$, respectively. (c) Measured intrinsic $Q$-factors of untargeted modes for different modulation amplitudes, showing consistent values at the million level. (d-g) The amount of mode splitting as a function of the geometric modulation amplitude $A$, in cases of (d) different ring widths, (e) targeted ($m = 600$, solid dots) and adjacent modes ($m = 600\pm1$, circles), (f) different target mode numbers $m$, and (g) a doubled resonator radius.

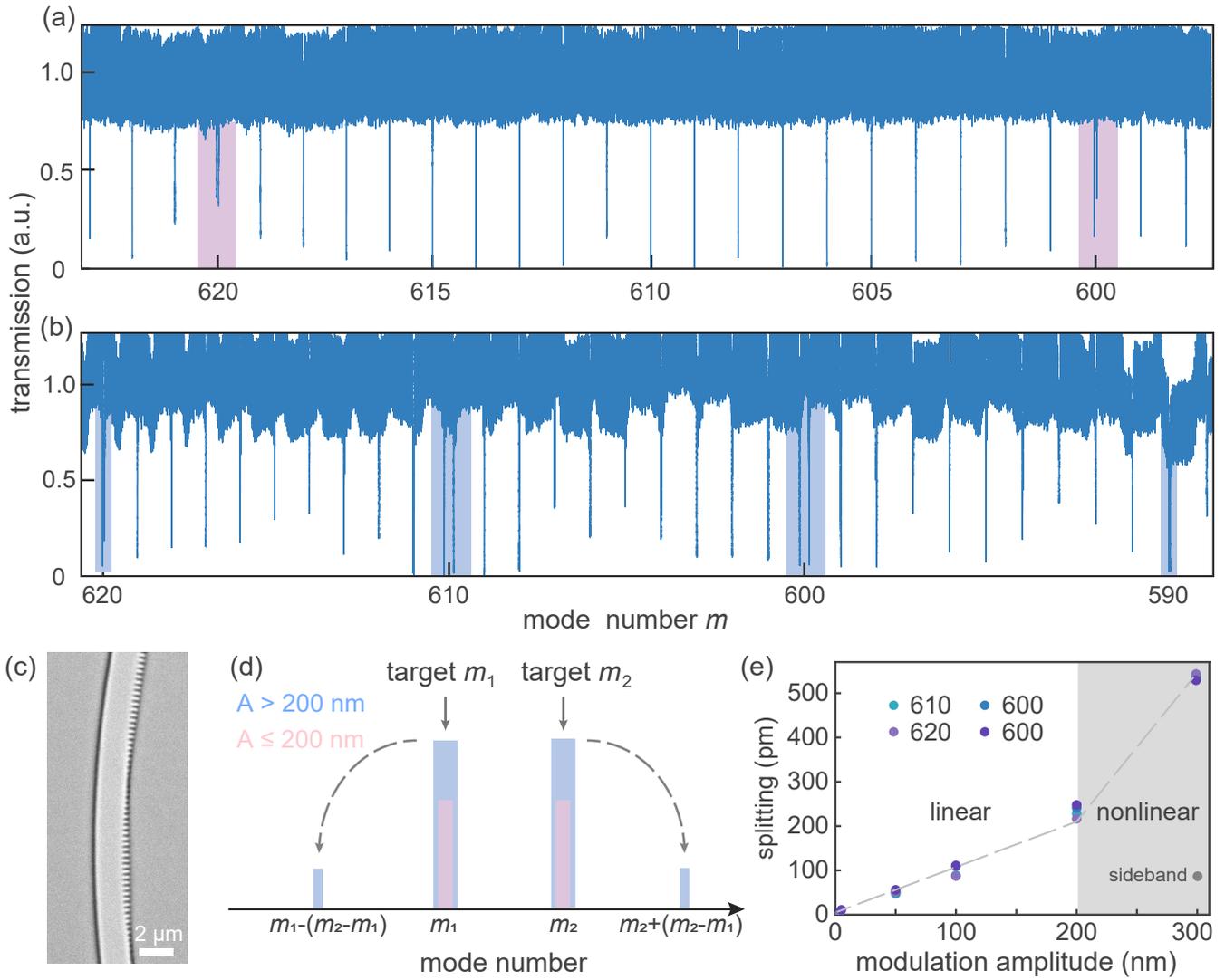

**Figure 3.** Dual-tone geometric modulation in z-cut LN PhCR resonators. (a) Optical transmission spectrum of a dual-tone moderately modulated z-cut LN PhCR ($A = 100$ nm), with two observable split modes 600 and 620 (pink shaded area). (b) Optical transmission spectrum of a dual-tone modulated z-cut LN PhCR in an aggressive region ($A = 300$ nm). Mode splitting occurs not only at target modes 600 and 610, but also at "sideband" modes 590 and 620. (c) SEM image of a dual-tone geometric modulated z-cut LN PhCR. (d) Schematic illustration of the generation of spurious mode splitting in an aggressively modulated case. (e) The amount of mode splitting as a function of modulation amplitude $A$ for various tested modes, showing a near-linear relationship in the moderate region ($A \leq 200$ nm) and a nonlinear relationship in the aggressive region ($A > 200$ nm, grey area). Grey dot represents the splitting amount of sideband modes at $A = 300$ nm.

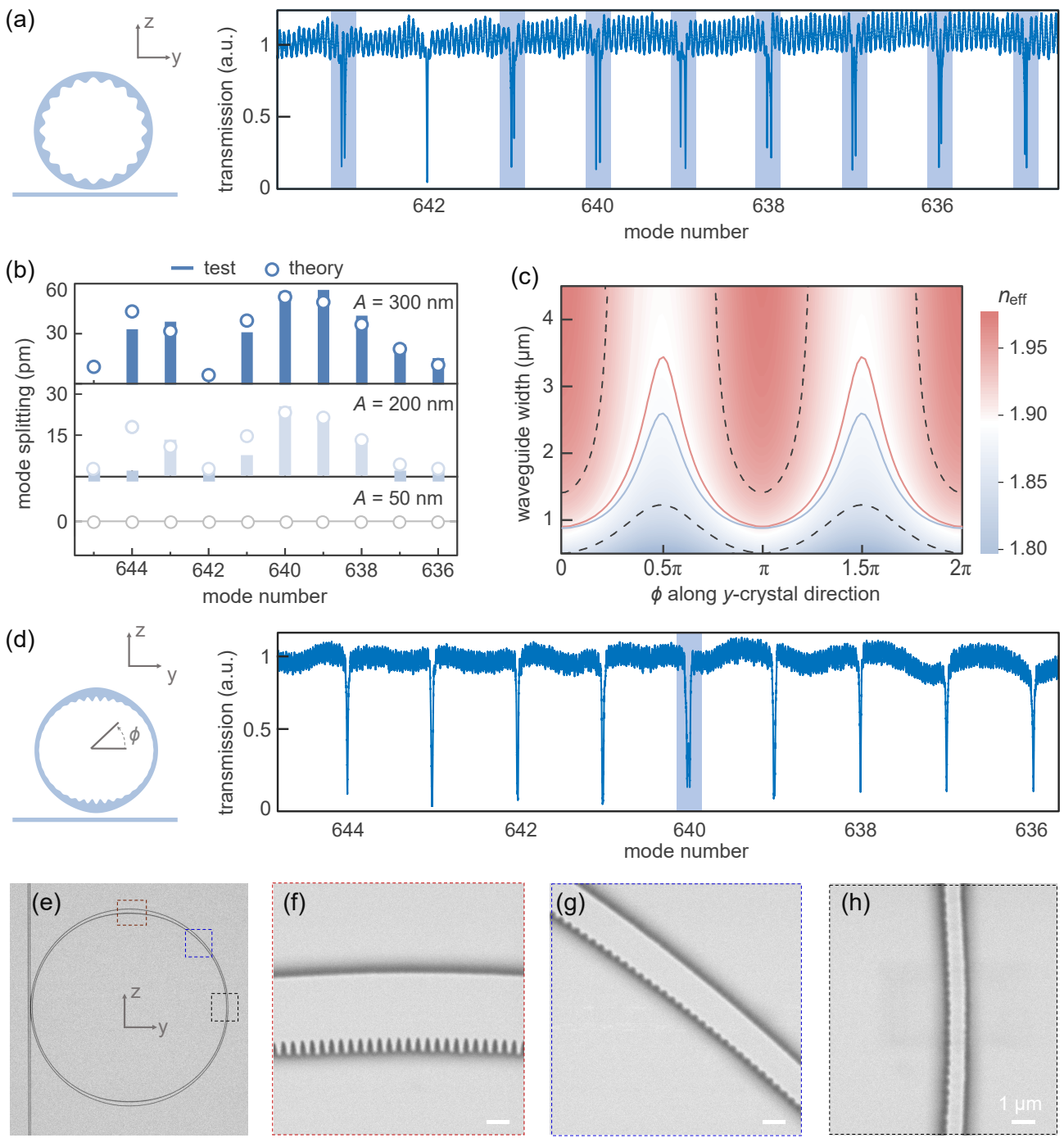

**Figure 4.** Gradient geometric modulation in x-cut LN PhCR resonators. (a) Schematic view (left) and measured optical transmission spectrum (right) of a simple periodic x-cut LN PhCR resonator, showing multiple split modes (shaded areas) around the target mode $m = 640$ due to the anisotropic nature of LN crystal. (b) Theoretically predicted (circles) and experimentally measured (bars) splitting amounts of 10 nearest modes around the target mode, for uniform modulation amplitudes $A$ of 300 nm (top), 200 nm (middle), and 50 nm (bottom). (c) Calculated effective indices as functions of azimuthal angle and waveguide width in x-cut LN. The final width profiles of the x-cut LN PhCR at the tip and indent points follow the red and blue solid curves, respectively. (d) Measured transmission spectrum of the gradient x-cut LN PhCR resonator (schematic shown on the left), showing single mode splitting. (e-h) SEM images of the fabricated gradient x-cut LN PhCR resonator, with zoom-in views illustrating the geometric modulation profiles at different locations.

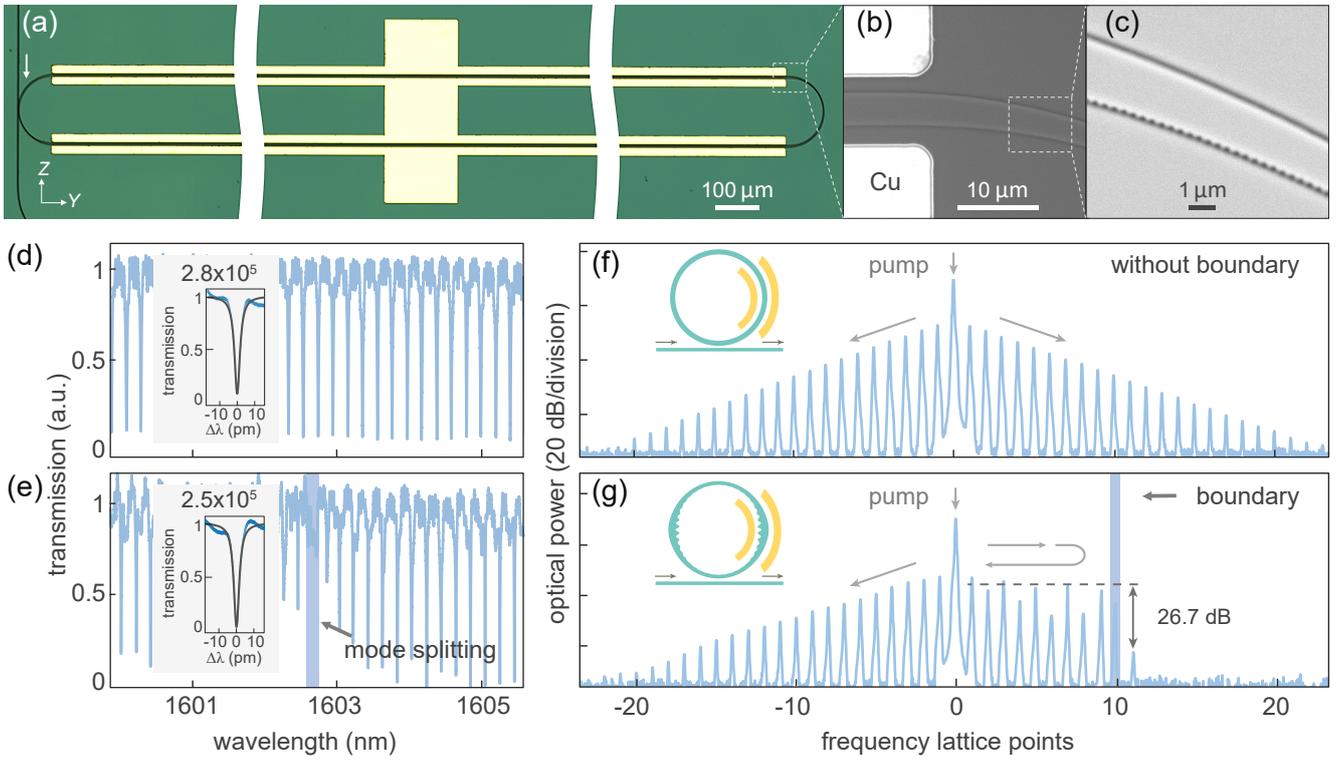

**Figure 5.** PhCR-induced boundary in the synthetic frequency dimension. (a) Microscope image of the fabricated device, consisting of a 24.96-GHz gradient PhCR resonator and modulation electrodes. (b-c) Zoom-in SEM images illustrating the EO modulation region (b) and the LN waveguide with a gradient geometric modulation profile (c), respectively. (d-e) Measured optical transmission spectra of a reference LN racetrack resonator (d) and our PhCR device (e), respectively. Insets show the zoom-in spectra of single resonances with Lorentzian fitting. (f-g) Measured frequency comb spectra showing one-way optical energy flow in a normal EO comb generator (f) and strong reflection at the mode-splitting-induced frequency boundary in our PhCR-based device (g). Insets show the schematic illustrations of the two devices. Shaded area in (g) corresponds to the mode-splitting location.